\documentclass[11pt]{article}
\usepackage{epsf,graphicx} 
\begin{document}

\begin{Large}
\begin{center}
\textbf{Role of antikaon condensation in r-mode instability}
\end{center}
\end{Large}
\begin{center}
Debarati Chatterjee and Debades Bandyopadhyay\\
\textit{Saha Institute of Nuclear Physics, 1/AF Bidhannagar,
Kolkata-700064, India}
\end{center}
\vspace{1cm}
\begin{abstract}
\indent We investigate the effect of antikaon condensed matter 
on bulk viscosity in rotating neutron stars. We use relativistic field 
theoretical models to construct the equation of state of neutron stars with the
condensate, where the phase transition from nucleonic to $K^-$ condensed phase 
is assumed to be of first order. We calculate the coefficient of bulk viscosity
due to the non-leptonic weak interaction $n \rightleftharpoons p + K^-$. The 
influence of antikaon bulk viscosity on the gravitational radiation reaction 
driven instability in the r-modes is investigated. We compare our results with 
the previously studied non-leptonic weak interaction 
$n + p \rightleftharpoons p + \Lambda$ involving hyperons on the damping of 
the r-mode oscillations.\\
\indent We find that the bulk viscosity coefficient due to the non-leptonic 
weak process involving the condensate is suppressed by several orders of 
magnitude in comparison with the non-superfluid hyperon bulk viscosity 
coefficient. Consequently, the antikaon bulk viscosity may not be able to damp 
the r-mode instability, while hyperon bulk viscosity can effectively suppress 
r-mode oscillations at low temperatures. Hence neutron stars containing $K^-$ 
condensate in their core could be possible sources of gravitational waves.
\end{abstract}
\section{Introduction}
\indent \indent Immense information about the internal composition of neutron 
stars can be obtained through the study of unstable modes of oscillations 
associated with rotating neutron stars. Among the various possible 
instabilities, inertial r-modes restored by Coriolis force are particularly 
interesting as they are thought to play an important role in regulating the 
spins of newly-born neutron stars as well as old, accreting neutron stars in 
low mass X-ray binaries (LMXBs). If the r-mode is unstable, a rapidly rotating 
neutron star could emit a significant fraction of its rotational energy and 
angular momentum as gravitational waves, which could be detectable by the 
upcoming generation of gravitational wave detectors.\\ 
\indent Bulk viscosity in neutron star is caused by the energy dissipation due 
to non-leptonic weak interaction processes in pulsating dense matter. It was 
argued that the r-mode instability could
be effectively suppressed by bulk viscosity due to exotic matter in neutron 
star interior. Neutron star matter spans a wide range of densities, from 
the density of iron nucleus at the surface of the star to several times 
normal nuclear matter density in the core. Since the chemical potentials of
nucleons and leptons increase rapidly with density in the neutron
star core, different exotic  forms of matter with large strangeness fraction such 
as hyperons, Bose-Einstein condensates of antikaons 
or deconfined quarks may appear there. The coefficient of bulk viscosity due to 
non-leptonic weak processes involving hyperons was calculated by several
authors \cite{Jon1,Jon2,Lin02,Dal,Dra}. Also, the impact of bulk viscosity due to unpaired and 
paired quark matter on the r-mode instability had been investigated extensively
\cite{Dra,Mad92,Mad00}.\\
\indent In this paper, we study the effect of the presence of antikaon condensates on the equation of state (EoS), bulk viscosity and the corresponding damping timescale. We compare the bulk viscosities generated by the non-leptonic weak interaction $n \rightleftharpoons p + K^-$ involving antikaons with that due to the non-leptonic process $n + p \rightleftharpoons p + \Lambda $ involving hyperons as previously obtained, and investigate their role in damping of r-mode oscillations in rotating neutron stars. \\
\indent This paper is 
organized in the following way. In Sec. II, we describe the model to calculate
equation of state, bulk viscosity coefficient and the corresponding 
timescale. The parameters of the model are listed in Sec. III. The results of our calculations are
discussed in Sec. IV, and the summary and conclusions are given in Sec. V. \\
\section{Theoretical Model}
We assume a first order phase transition from nuclear to antikaon condensed matter. A relativistic field theoretical model is adopted to describe the $\beta$-equilibrated and charge neutral matter in both the phases. \\
\subsection{Hadronic Phase}
The constituents of hadronic phase are octet baryons, electrons and muons. In this model, baryon-baryon interaction is mediated by the exchange
of scalar and vector mesons. For hyperon-hyperon interaction, two additional strange mesons, scalar $f_0$ (denoted by $\sigma^*$) and vector $\phi$ are incorporated.
The Lagrangian density for the hadronic phase is given by
\begin{eqnarray}
{\cal L}_B &=& \sum_B \bar\Psi_{B}\left(i\gamma_\mu{\partial^\mu} - m_B
+ g_{\sigma B} \sigma - g_{\omega B} \gamma_\mu \omega^\mu
- g_{\rho B}
\gamma_\mu{\mbox{\boldmath t}}_B \cdot
{\mbox{\boldmath $\rho$}}^\mu \right)\Psi_B\nonumber\\
&& + \frac{1}{2}\left( \partial_\mu \sigma\partial^\mu \sigma
- m_\sigma^2 \sigma^2\right) - U(\sigma) \nonumber\\
&& -\frac{1}{4} \omega_{\mu\nu}\omega^{\mu\nu}
+\frac{1}{2}m_\omega^2 \omega_\mu \omega^\mu
- \frac{1}{4}{\mbox {\boldmath $\rho$}}_{\mu\nu} \cdot
{\mbox {\boldmath $\rho$}}^{\mu\nu}
+ \frac{1}{2}m_\rho^2 {\mbox {\boldmath $\rho$}}_\mu \cdot
{\mbox {\boldmath $\rho$}}^\mu + {\cal L}_{YY} .
\end{eqnarray}
The isospin multiplets for baryons B are
represented by the Dirac bispinor $\Psi_B$ with vacuum baryon mass $m_B$,
and isospin operator ${\mbox {\boldmath t}}_B$, and $\omega_{\mu\nu}$ and 
$\rho_{\mu\nu}$ are field strength tensors. The scalar
self-interaction term \cite{Bog} 
\begin{equation}
U(\sigma) = \frac{1}{3} g_2 \sigma^3 + \frac{1}{4} g_3 \sigma^4 ,
\end{equation}
is introduced to reproduce the correct compressibility of nuclear matter.
We perform this calculation in the mean field approximation \cite{Ser}. The Lagrangian density for hyperon-hyperon interaction is given by
\begin{eqnarray}
{\cal L}_{YY} &=& \sum_B \bar\Psi_{B}\left(
 g_{\sigma^* B} \sigma^* - g_{\phi B} \gamma_\mu \phi^\mu
 \right)\Psi_B\nonumber\\
&& + \frac{1}{2}\left( \partial_\mu \sigma^* \partial^\mu \sigma^*
- {m_\sigma^*}^2 {\sigma^*}^2\right) -\frac{1}{4} \phi_{\mu\nu}\phi^{\mu\nu}
+ \frac{1}{2} m_\phi^2 \phi_\mu \phi^\mu.
\end{eqnarray} 
The scalar density and baryon number density are 
\begin{eqnarray}
n_B^S &=& \frac{2J_B+1}{2\pi^2} \int_0^{k_{F_B}} 
\frac{m_B^*}{(k^2 + m_B^{* 2})^{1/2}} k^2 \ dk ~,
\end{eqnarray}
\begin{eqnarray}
n_B &=& (2J_B+1)\frac{k^3_{F_B}}{6\pi^2} ~, 
\end{eqnarray}
where Fermi momentum is $k_{F_B}$, spin is $J_B$, and isospin projection is
$I_{3B}$. Effective mass and chemical potential of baryon $B$ are 
$m_B^*=m_B - g_{\sigma B}\sigma - g_{\sigma^* B}\sigma^*$ and
$\mu_{B} = (k^2_{F_{B}} + m_B^{* 2} )^{1/2} + g_{\omega B} \omega_0
+ g_{\phi B} \phi_0 + I_{3B} g_{\rho B} \rho_{03}$, respectively. Charge neutrality in the hadronic phase is imposed through the condition
\begin{equation}
Q = \sum_B q_B n_B -n_e -n_\mu =0~,
\end{equation}
where $n_B$ is the number density of baryon B, $q_B$ is the electric charge 
and $n_e$ and $ n_\mu$ are charge densities of electrons and muons respectively.
The total energy density in the hadronic phase is given by 
\begin{eqnarray}
{\varepsilon}
&=& \frac{1}{2}m_\sigma^2 \sigma^2 
+ \frac{1}{3} g_2 \sigma^3 + \frac{1}{4} g_3 \sigma^4
+ \frac{1}{2}m_{\sigma^*}^2 \sigma^{*2}\nonumber\\ 
&& + \frac{1}{2} m_\omega^2 \omega_0^2 + \frac{1}{2} m_\phi^2 \phi_0^2 
+ \frac{1}{2} m_\rho^2 \rho_{03}^2 \nonumber\\
&& + \sum_B \frac{2J_B+1}{2\pi^2} 
\int_0^{k_{F_B}} (k^2+m^{* 2}_B)^{1/2} k^2 \ dk \nonumber\\
&& + \sum_{l=e^-,\mu^-} \frac{1}{\pi^2} \int_0^{K_{F_l}} (k^2+m^2_l)^{1/2} k^2 \ dk,
\end{eqnarray}
and the pressure is 
\begin{eqnarray}
P &=& - \frac{1}{2}m_\sigma^2 \sigma^2 - \frac{1}{3} g_2 \sigma^3 
- \frac{1}{4} g_3 \sigma^4 \nonumber\\
&& - \frac{1}{2}m_{\sigma^*}^2 \sigma^{*2} 
+ \frac{1}{2} m_\omega^2 \omega_0^2 + \frac{1}{2} m_\phi^2 \phi_0^2 
+ \frac{1}{2} m_\rho^2 \rho_{03}^2 \nonumber\\
&& + \frac{1}{3}\sum_B \frac{2J_B+1}{2\pi^2} 
\int_0^{k_{F_B}} \frac{k^4 \ dk}{(k^2+m^{* 2}_B)^{1/2}}\nonumber\\
&& + \frac{1}{3} \sum_{l=e^-,\mu^-} \frac{1}{\pi^2} 
\int_0^{K_{F_l}} \frac{k^4 \ dk}{(k^2+m^2_l)^{1/2}}~. 
\end {eqnarray}
\subsection{Antikaon condensed phase}
\indent The constituents of the pure antikaon condensed phase are baryons (neutrons, protons), leptons (electrons, muons) and antikaons, where the baryons are embedded in the condensate. The (anti)kaon-(anti)kaon interaction in the pure condensed phase is described using the relativistic field theoretical approach, through the exchange of $\sigma$, $\omega$, $\rho$, $\sigma^*$ and $\phi$ mesons. However, nucleons do not couple with the strange mesons, hence $g_{\sigma^* N} = g_{\phi N} = 0$. \\
\indent The Lagrangian density for (anti)kaons in the minimal coupling scheme is, 
\begin{equation}
{\cal L}_K = D^*_\mu{\bar K} D^\mu K - m_K^{* 2} {\bar K} K ~,
\end{equation}
where the covariant derivative
$D_\mu = \partial_\mu + ig_{\omega K}{\omega_\mu} + ig_{\phi K}{\phi_\mu}
+ i g_{\rho K}
{\mbox{\boldmath t}}_K \cdot {\mbox{\boldmath $\rho$}}_\mu$ and
the effective mass of (anti)kaons is 
$m_K^* = m_K - g_{\sigma K} \sigma - g_{\sigma^* K} \sigma^*$. The in-medium energies of
$K^-$ mesons for $s$-wave (${\vec k}=0$) condensation is given by
\begin{equation}
\mu_{K^-} = m_K^* - g_{\omega K} \omega_0 - g_{\phi K} \phi_0
- \frac{1}{2} g_{\rho K} \rho_{03} ~,
\end{equation}
where the isospin projection $I_{3K^-} = -1/2$. The scalar and number density of baryon $B$ in the antikaon condensed 
phase are given by
\begin{eqnarray}
n_B^{{K},S} &=& \frac{2J_B+1}{2\pi^2} \int_0^{k_{F_B}}
\frac{m_B^*}{(k^2 + m_B^{* 2})^{1/2}} k^2 \ dk ~,\\
n_B^{K} &=& (2J_B+1)\frac{k^3_{F_B}}{6\pi^2} ~,
\end{eqnarray}
\indent The scalar density of $K^-$ mesons in the condensate is given 
by \cite{Gle99}
\begin{equation}
n_{K^-} = 2\left( \omega_{K^-} + g_{\omega K} \omega_0 
+ g_{\phi K} \phi_0 + \frac{1}{2} g_{\rho K} \rho_{03} \right) {\bar K} K  
= 2m^*_K {\bar K} K  ~.
\end{equation}

The total charge density in the antikaon condensed phase is 
\begin{equation}
Q^{K}=\sum_{B=n,p} q_B n_B^{K} - n_{K^-} - n_e - n_\mu =0.
\end{equation}

The total energy density in the antikaon condensed phase is
\begin{eqnarray}
{\varepsilon^{K}}  &=& \frac{1}{2}m_\sigma^2 \sigma^2
+ \frac{1}{3} g_2 \sigma^3
+ \frac{1}{4} g_3 \sigma^4  + \frac{1}{2}m_{\sigma^*}^2 \sigma^{*2}
+ \frac{1}{2} m_\omega^2 \omega_0^2 + \frac{1}{2} m_\phi^2 \phi_0^2
+ \frac{1}{2} m_\rho^2 \rho_{03}^2  \nonumber \\
&& + \sum_{B=n,p} \frac{2J_B+1}{2\pi^2}
\int_0^{k_{F_B}} (k^2+m^{* 2}_B)^{1/2} k^2 \ dk
+ \sum_{l=e^-,\mu^-} \frac{1}{\pi^2} \int_0^{K_{F_l}} (k^2+m^2_l)^{1/2} k^2 \ dk
\nonumber \\
 && + m^*_K n_{K^-},
\end{eqnarray}
where last term denotes the contribution of the $K^-$ condensate.
The pressure in this phase is
\begin{eqnarray}
P^{K} &=& - \frac{1}{2}m_\sigma^2 \sigma^2 - \frac{1}{3} g_2 \sigma^3
- \frac{1}{4} g_3 \sigma^4  - \frac{1}{2}m_{\sigma^*}^2 \sigma^{*2}
+ \frac{1}{2} m_\omega^2 \omega_0^2 + \frac{1}{2} m_\phi^2 \phi_0^2
+ \frac{1}{2} m_\rho^2 \rho_{03}^2 \nonumber \\
&& + \frac{1}{3}\sum_{B=n,p} \frac{2J_B+1}{2\pi^2}
\int_0^{k_{F_B}} \frac{k^4 \ dk}{(k^2+m^{* 2}_B)^{1/2}}
+ \frac{1}{3} \sum_{l=e^-,\mu^-} \frac{1}{\pi^2}
\int_0^{K_{F_l}} \frac{k^4 \ dk}{(k^2+m^2_l)^{1/2}}~.
\end{eqnarray}
\subsection{The Mixed Phase}

\indent \indent The mixed phase of hadronic and $K^-$ condensed matter is governed by the 
Gibbs phase equilibrium rules \cite{Gle99,Gle92},
\begin{eqnarray}
P^h&=& P^{K},\\
\mu_B^h& =& \mu_B^{K}.
\end{eqnarray}
where $\mu_B^h$ and $\mu_B^{K}$ are chemical potentials of baryons B in the
pure hadronic and $K^-$ condensed phase, respectively.
The global charge neutrality and baryon number conservation laws are 
\begin{equation}
(1-\chi) Q^h + \chi Q^{K} = 0,\\
\end{equation}
\begin{equation}
n_B=(1-\chi) n_B^h + \chi n_B^{K}~,
\end{equation}
where $\chi$ is the volume fraction of $K^-$ condensed phase in the mixed 
phase. The total energy density in the mixed phase is given by 
\begin{equation}
\epsilon=(1-\chi)\epsilon^h + \chi \epsilon^{K}~.
\end{equation}

\subsection{Bulk Viscosity}
\indent \indent Energy dissipation due to pressure and density variations associated with r-mode oscillation, which drive the system out of $\beta$ equilibrium, gives rise to bulk viscosity. The reactions between different constituent particles try to bring the system back to an equilibrium configuration, with a delay which depends on the characteristic timescale of the interaction. Strong interaction processes are insignificant as the strong interaction equilibrium is reached so fast that these processes are considered to be in equilibrium compared to typical pulsation timescales.\\
\indent As we are concerned about bulk viscosity coefficient in young neutron stars where 
temperature is $\sim$ 10$^9$ - 10$^{10}$ K, we want to find out whether non-leptonic processes involving antikaons might lead to a high value for the bulk viscosity coefficient. The relevant non-leptonic process involving antikaons is 
\begin{equation}
n \rightleftharpoons p + K^-.
\end{equation}

As this reaction involves variation of neutron number density ($n_n$) due to 
density perturbation, we consider neutron fraction
as a primary parameter.
The general expression for the real part of bulk viscosity coefficient \cite{Lin02,Lan} is
\begin{equation}
Re \> \zeta = \frac {P(\gamma_{\infty} - \gamma_0)\tau}{1 + {(\omega\tau)}^2}~,
\end{equation}
where $P$ is the pressure, $\tau$ is the net microscopic relaxation time and 
$\gamma_{\infty}$ and $\gamma_0$ are 'infinite' and 'zero' frequency adiabatic 
indices respectively. The factor
\begin{equation}
\gamma_{\infty} - \gamma_0 = - \frac {n_b^2}{P} \frac {\partial P} 
{\partial n_n} \frac {d{\bar x}_n} {dn_b}~,
\end{equation}
can be determined from the EoS. Here $\bar x_n = \frac {n_n}{n_b}$ gives the 
neutron fraction in the equilibrium state and $n_b = {\sum}_{B} n_B$ is the 
total baryon density. In the co-rotating frame, the generic equation relating the angular velocity 
($\omega$) of $(l,m)$ r-mode to the angular velocity of rotation of the star ($\Omega$) is $\omega = {\frac {2m}{l(l+1)}} \Omega$ \cite{And01}.\\
\\
The relaxation time ($\tau$) for the process is given 
by \cite{Lin02}
\begin{equation}
\frac {1}{\tau} = \frac{\Gamma_K}{\delta \mu} \frac{\delta \mu}{\delta n_n^K}.
\end{equation}
Here $\delta {n_n^K} = n_n^K - {\bar n}_n^K$ is the departure of neutron fraction
from its thermodynamic equilibrium value ${\bar n}_n^K$ in the $K^-$ condensed phase. 
The reaction rate per unit volume is \cite{Rati2}
\begin{equation}
\Gamma_K = \frac{<|M_K|^2>  k_{F_n}^2 \delta \mu}{16 \pi^3 \mu_{K^-}},
\end{equation}
where $k_{F_n}$ is the Fermi momentum for neutrons in the condensed phase and 
$<{|M_K|}^2>$ is the squared matrix element, averaged over initial spins and summed over final spins. The in-medium energy of $K^-$ mesons in the condensate is $E_3 = \mu_3 = \mu_{K^-}$. If the quantity 
$\frac {\delta \mu}{\delta{n_n^K}}$ is calculated numerically, as soon as we know the
relaxation time, we can calculate the bulk viscosity coefficient.

\subsection{Matrix element}

Next, we focus on the evaluation of the matrix element for the non-leptonic process (22). In general, the matrix element for the decay of a $\frac{1}{2}^+$ baryon to another $\frac{1}{2}^+$ baryon and a $0^-$ meson can be written as
\begin{equation}
{\cal{M}} = \bar u (k_2) ( A + B \gamma_5 ) u(k_1)
\end{equation}
where $u(k_1)$ and $u(k_2)$ are the spinors of neutrons and protons respectively. Here A is the parity-violating amplitude, and B is the parity-conserving amplitude. The squared and spin averaged matrix element is given by
\begin{equation}
<{\cal{|M|}}^2> = 2 [(k_1 \cdot k_2 + m_n^* m_p^* ) |A|^2 
+ ( k_1 \cdot k_2 - m_n^* m_p^* ) |B|^2].
\end{equation} 
For $s$-wave $K^-$ condensation, $\vec k_3 = $0 $\Longrightarrow$  $|\vec k_1| = 
|\vec k_2|$. As fermion momenta lie close to the Fermi surfaces, 
$k_1 \cdot k_2 = E_1 E_2 - |\vec k_1||\vec k_2| \cos \theta 
= \mu_n \mu_p - k_{F_n} k_{F_p}$. This leads to the squared matrix element
\begin{equation}
<|\mathcal{M}^2|> = 2  [(\mu_n \mu_p - k_{F_n} k_{F_p} + m_n^* m_p^*)|A|^2 
+ (\mu_n \mu_p - k_{F_n} k_{F_p} - m_n^* m_p^*)|B|^2 ].
\end{equation}

Here we apply the weak SU(3) symmetry to the non-leptonic weak decay amplitudes
for the process (22). The weak decays of the octet hyperons can be
described by an effective SU(3) interaction with a parity violating (A) and 
parity conserving (B) amplitudes \cite{Mar,Sch00}. The weak operator is 
proportional to Gell-Mann matrix $\lambda_6$ to ensure hypercharge
violation $|\Delta Y| = 1$ and $|\Delta I| = 1/2$. Similarly, the amplitudes for (22) are extracted from experimentally known decay parameters of the weak decay of hyperons
\cite{Sch00}. The amplitudes are $A = - 1.62 \times 10^{-7}$ and 
$B = -7.1 \times 10^{-7}$. It is to be noted that all quantities in (29) are to be calculated in the condensed phase.\\
\subsection{Critical Angular Velocity}

The bulk viscosity damping timescale ($\tau_B$) due to the non-leptonic process 
involving antikaons is given by \cite{Nar,Lin02} 
\begin{equation}
{\frac {1}{\tau_B}} =  - {\frac {1} {2E}} {\frac {dE}{dt}}~,
\end{equation}
where E is the energy of the perturbation as measured in the co-rotating frame
of the fluid and is expressed as
\begin{equation}
E = \frac {1}{2}{\alpha^2}{\Omega^2}{R^{-2}} \int_0^R {\epsilon (r) r^6}dr~.
\end{equation}
Here, $\alpha$ is the dimensionless amplitude of the r-mode, R is the 
radius of the star and $\epsilon(r)$ is the energy density profile. The 
derivative of the co-rotating frame energy with respect
to time is
\begin{equation}
\frac {dE}{dt} = -4 \pi \int_0^R \zeta (r) <|\vec{\nabla} \cdot 
{\delta \vec{v}}|^2> r^2 dr,~
\end{equation} 
where the angle average of the square of the hydrodynamic expansion 
\cite{Lin99} is 
$$<|\vec{\nabla} \cdot {\delta \vec{v}}|^2> = \frac 
{({\alpha \Omega})^2}{690}
\left({\frac {r}{R}} \right)^6 \left(1 + 0.86 \left({\frac {r}{R}} \right)^2 \right) \left({\frac {\Omega^2} {\pi G \bar {\epsilon}}} \right)^2,$$
and $\bar {\epsilon}$ is the mean energy density of a non-rotating star. 
 The total r-mode time scale ($\tau_r$) is defined
as
\begin{equation}
{\frac {1}{\tau_r}} =  - {\frac {1}{\tau_{GR}}} + {\frac {1}{\tau_B}} + 
{\frac {1}{\tau_U}}.~ 
\end{equation}
where the time scales for 
gravitational radiation ($\tau_{GR}$) and modified Urca process ($\tau_U$) 
involving only nucleons have also been included. 
The gravitational radiation timescale is given by \cite{Lin98}
\begin{equation}
{\frac {1}{\tau_{GR}}} =  
\frac {131072 \pi}{164025} {\Omega^6} \int_0^R {\epsilon (r) r^6}dr~.
\end{equation}
The time scale due to modified Urca process ($\tau_U$) involving only nucleons 
is calculated from (31)
using the following expression for bulk viscosity coefficient for modified Urca 
process \cite{Lin98,Saw}
\begin{equation}
\zeta_U = 6 \times 10^{-59} \epsilon^2 T^6 \omega^2~.
\end{equation}
For a star of given mass, solving $\frac {1}{\tau_r}$ = 0, we can obtain the critical 
angular velocity at each temperature above which the r-mode becomes unstable. 

\section{Parameters of the Theory}

\subsection{Nucleon-Meson coupling constants}
Nucleon-meson coupling constants are determined from saturation properties of nuclear matter \cite{Gle91}. We used the following values: binding energy $ =-16.3$ MeV, baryon density $n_0=0.153$ fm$^{-3}$, asymmetry energy coefficient $a_{\rm asy}=32.5$ MeV, incompressibility $K=240$ MeV, and effective nucleon mass $m^*_N/m_N = 0.78$. We also studied the parameter set \cite{Bani2} with
incompressibility $K=300$ MeV, and effective nucleon mass $m^*_N/m_N = 0.70$.

\subsection{Kaon-Meson coupling constants}
According to the quark model and isospin counting rule, the vector coupling constants are
given by
\begin{equation}
g_{\omega K} = \frac{1}{3} g_{\omega N} ~~~~~ {\rm and} ~~~~~
g_{\rho K} = g_{\rho N} ~.
\end{equation}
The scalar coupling constant is obtained from the real part of
$K^-$ optical potential depth at normal nuclear matter density
\begin{equation}
U_{\bar K} \left(n_0\right) = - g_{\sigma K}\sigma - g_{\omega K}\omega_0 ~.
\end{equation}
It is known from $K^-$-atomic data that antikaons experience an 
attractive potential in nuclear matter while kaons feel a repulsive interaction 
\cite{Fri94,Fri99,Koc,Waa,Li,Pal2}. The strength of antikaon optical potential depth
ranges from shallow attractive ($-40$ MeV) to strongly attractive ($-180$ MeV).
Here we perform the calculation for antikaon optical potential depth at normal nuclear matter 
density $U_{\bar K}(n_0) = -120$ MeV.
The strange meson fields couple with (anti)kaons.
The $\sigma^*$-K coupling constant is $g_{\sigma^*K}=2.65$ as 
determined from the decay of $f_0$(925) meson and the vector $\phi$ meson
coupling with (anti)kaons $\sqrt{2} g_{\phi K} = 6.04$ follows from
the SU(3) relation \cite{Sch}. \\

\subsection{Hyperon-Meson coupling constants}
Hyperon-vector meson coupling constants are determined from SU(6) symmetry of
the quark model \cite{Mis,Dov,Sch94}. The scalar $\sigma$ meson 
coupling to hyperons is calculated from the potential depth of a hyperon (Y) 
\begin{equation}
U_Y^N (n_0) = - g_{\sigma Y} \sigma + g_{\omega Y} \omega_0~,
\end{equation}
in normal nuclear matter. The potential depth of $\Lambda$ hyperons in normal
nuclear matter $U_{\Lambda}^N (n_0) = -30$ MeV is obtained from the analysis
of energy levels of $\Lambda$ hypernuclei \cite{Dov,Chr}. Recent 
$\Xi$-hypernuclei data from various experiments \cite{Fuk,Kha} give 
a relativistic potential of $U_{\Xi}^N (n_0) = -18$ MeV. However, the analysis 
of $\Sigma^-$ 
atomic data implies a strong isoscalar repulsion for $\Sigma^-$ hyperons in
nuclear matter \cite{Fri}. Also, recent $\Sigma$ hypernuclei data indicate a 
repulsive
$\Sigma$-nucleus potential depth \cite{Bart}. Therefore, a repulsive
potential depth of 30 MeV for $\Sigma$ hyperons \cite{Fri} is adapted. 

The hyperon-$\sigma^*$ coupling constants are estimated by fitting them to a
potential depth, ${U_{Y}^{(Y^{'})}}{(n_0)}$, for a hyperon (Y) in a hyperon 
($Y^{'}$) matter at normal nuclear matter density obtained from double 
$\Lambda$ hypernuclei data \cite{Sch93,Mis}. This is given by
\begin{equation}
U_{\Xi}^{(\Xi)}(n_0) = U_{\Lambda}^{(\Xi)}(n_0) = 2 U_{\Xi}^{(\Lambda)}(n_0)
= 2 U_{\Lambda}^{(\Lambda)}(n_0) = -40~.
\end{equation}

\section{Results and Discussions}
\indent \indent The Equation of State (pressure versus energy density) for neutron star matter with $K^-$ condensate is plotted in Fig. 1 (solid line) for K=240 MeV. The equations of state for neutron star matter containing nucleons only (short dashed line) and with hyperons (long dashed line) are superimposed on the same figure. The two kinks on the EoS involving the condensate at 3.26 $n_0$ and 4.62 $n_0$ mark the beginning and end of the mixed phase. For $K^-$ condensed matter with K=300 MeV, similar kinks are observed at 2.23 $n_0$ and 3.59 $n_0$ defining the mixed phase. The threshold density for the appearance of the $\Lambda$ hyperon is 2.6 $n_0$. The EoS becomes softer in presence of exotic matter ($K^-$ condensate or hyperons) compared with that of nucleon matter. However, the EoS for antikaon condensed matter is stiffer than that for hyperon matter beyond the mixed phase. \\ 
\indent For the calculation of damping time scale due to bulk viscosity using (30), we need the energy density profile and bulk viscosity profile of the neutron star. We choose a neutron star of gravitational mass 1.63$M_{\odot}$ corresponding to a central baryon density 3.94 $n_0$ and rotating at an angular velocity $\Omega = 1180 s^{-1}$. The neutron star is so chosen to ensure that it contains $K^-$ condensate in its core because the central baryon density is well above the 
threshold of $K^-$ condensation (3.26 $n_0$). The bulk viscosity profiles for K = 240 MeV and K = 300 MeV are plotted in Fig 2 as a function of equatorial distance. Here we note that the bulk viscosity profile drops to zero value beyond a certain equatorial distance, when the baryon density decreases below the threshold density of $K^-$ condensation and the non-leptonic process in (22) ceases to occur in the star.\\
\indent We compare the bulk viscosity coefficient due to antikaon condensate with bulk viscosity due to the non-leptonic weak interaction $n + p \rightleftharpoons p + \Lambda$ involving hyperons. The coefficient of bulk viscosity due to non-leptonic process involving hyperons is shown in Fig 3 as a function of normalised baryon density for different temperatures \cite{Rati}. Comparing the two figures 2 and 3, we can infer that the bulk viscosity coefficient in antikaon condensed matter is suppressed by several orders of magnitude in comparison with nonsuperfluid hyperon bulk viscosity. It must also be noted that the antikaon bulk viscosity is independent of temperature, while hyperon bulk viscosity increases with decrease in temperature. Hence, hyperon bulk viscosity can act as an effective damping mechanism at low temperatures.\\
\indent We can obtain critical angular velocities as a function of temperature by solving $\frac{1}{\tau_r} = 0$ in (33) for a rotating neutron star of mass 1.63 $M_{solar}$. We calculate the bulk viscosity coefficient due to modified Urca process due to nucleons using (35). The modified Urca bulk viscosity is plotted as a function of normalised baryon density for a range of temperatures in Fig 5. It is evident from the figure that bulk viscosity due to modified Urca process involving nucleons increases with increase in temperature. Hence r-mode instability is effectively suppressed by modified Urca bulk viscosity at high temperatures. The ratio of critical angular velocities to the rotational velocity of the neutron star are plotted as a function of temperature in Fig 4 for non-leptonic processes involving antikaons as well as hyperons. In this figure, we observe that there exists a window of instability for matter with hyperons. We can interpret that hyperon bulk viscosity damps the r-mode instability at low temperatures, whereas it is effectively suppressed at high temperatures by the bulk viscosity due to the modified Urca process involving nucleons. For $K^-$ condensed matter, we can infer that r-mode instability is damped by nucleonic modified Urca process at high temperatures. From the figure, it is evident that antikaon bulk viscosity is not an effective mechanism to damp the instability at low temperatures. The argument is further justified by the fact that the critical velocity curves obtained for both parameter sets K= 240 MeV and K = 300 MeV are dictated purely by the bulk viscosity due to nucleonic modified Urca process. The bulk viscosity due to antikaon condensation is not sufficient to suppress the r-mode within the temperature range considered here. Hence the instability window is wider for $K^-$ condensed matter, and neutron stars with $K^-$ condensate in its interior could be possible sources of gravitational waves.
\section{Summary} 
The role of $K^-$ condensation on bulk viscosity and 
r-mode instability has been investigated in this paper. We have estimated the bulk viscosity coefficient and the corresponding damping time
scale due to the non-leptonic process $n \rightleftharpoons p + K^{-}$ 
and compared them with those associated with the non-leptonic process $n + p \rightleftharpoons p + \Lambda$ involving hyperons. We have considered a first order phase transition
from the nuclear to the antikaon condensed phase, and the equation of state has been constructed using relativistic mean field theoretical models. We find that the bulk
viscosity coefficient in $K^-$ condensed is suppressed by several orders of magnitude in comparison with the nonsuperfluid hyperon bulk viscosity \cite{Lin02,Nar,Rati}. We also infer
that the antikaon bulk viscosity is unable to damp the r-mode
instability unlike hyperon bulk viscosity. If the instability is not suppressed by the viscosity, the star is forced to lose angular momentum as it cools down via gravitational radiation. So, r-modes in neutron stars with antikaon condensate in their core are possible candidates of gravitational waves.\\ 
{
  
} 
\newpage
\vspace{-2cm}

{\centerline{
\epsfxsize=12cm
\epsfysize=14cm
\epsffile{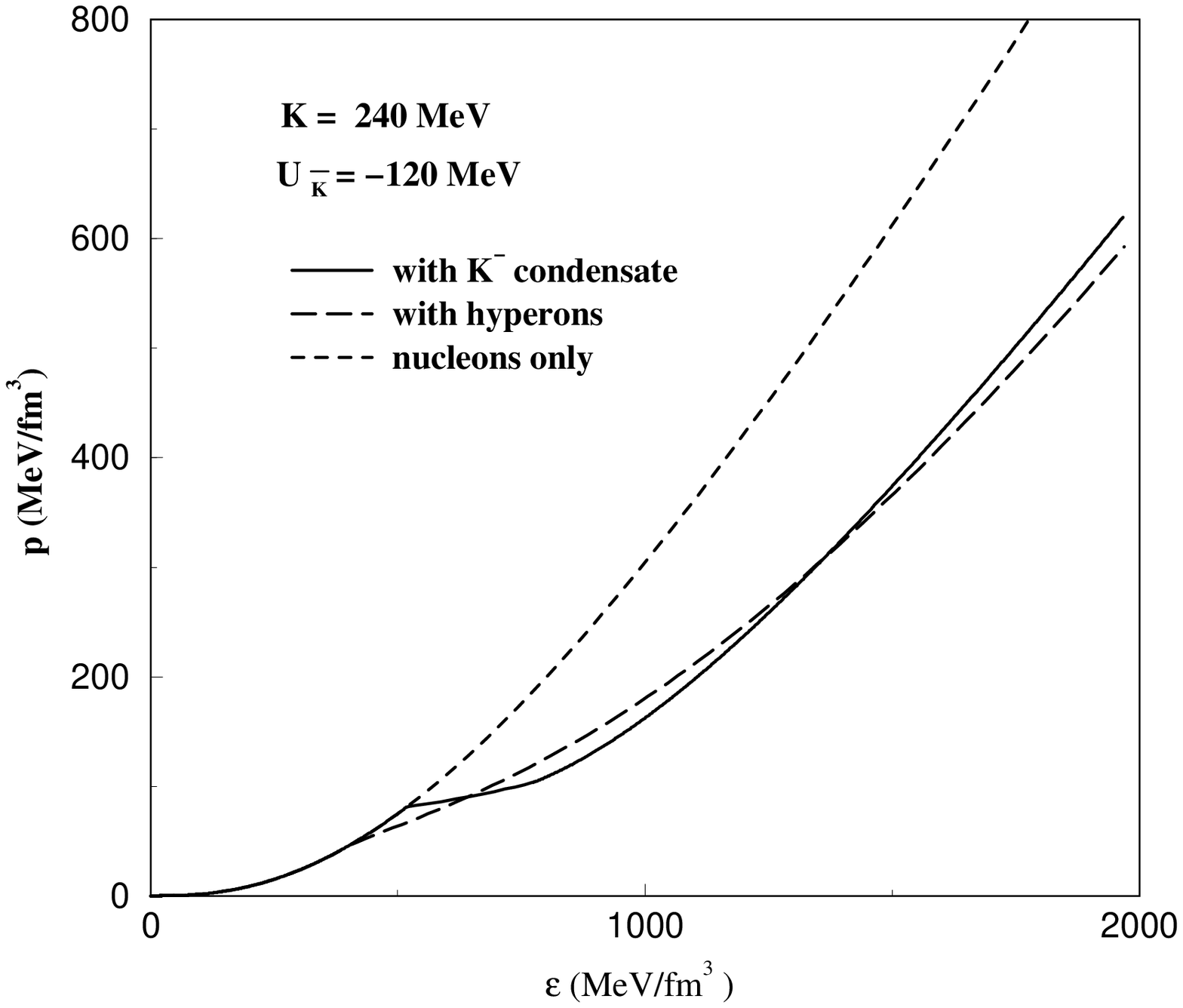}
}}

\vspace{4.0cm}

\noindent{\small{
Fig. 1. The equation of state ( pressure P vs energy density $\epsilon$ ) for
matter containing nucleons only (short dashed line), with hyperons (long dashed line) and with $K^-$ condensate (solid line) for antikaon optical potential depth at normal nuclear 
matter density $U_{\bar K} = -120$ MeV.}}

\newpage
\vspace{-2cm}

{\centerline{
\epsfxsize=12cm
\epsfysize=14cm
\epsffile{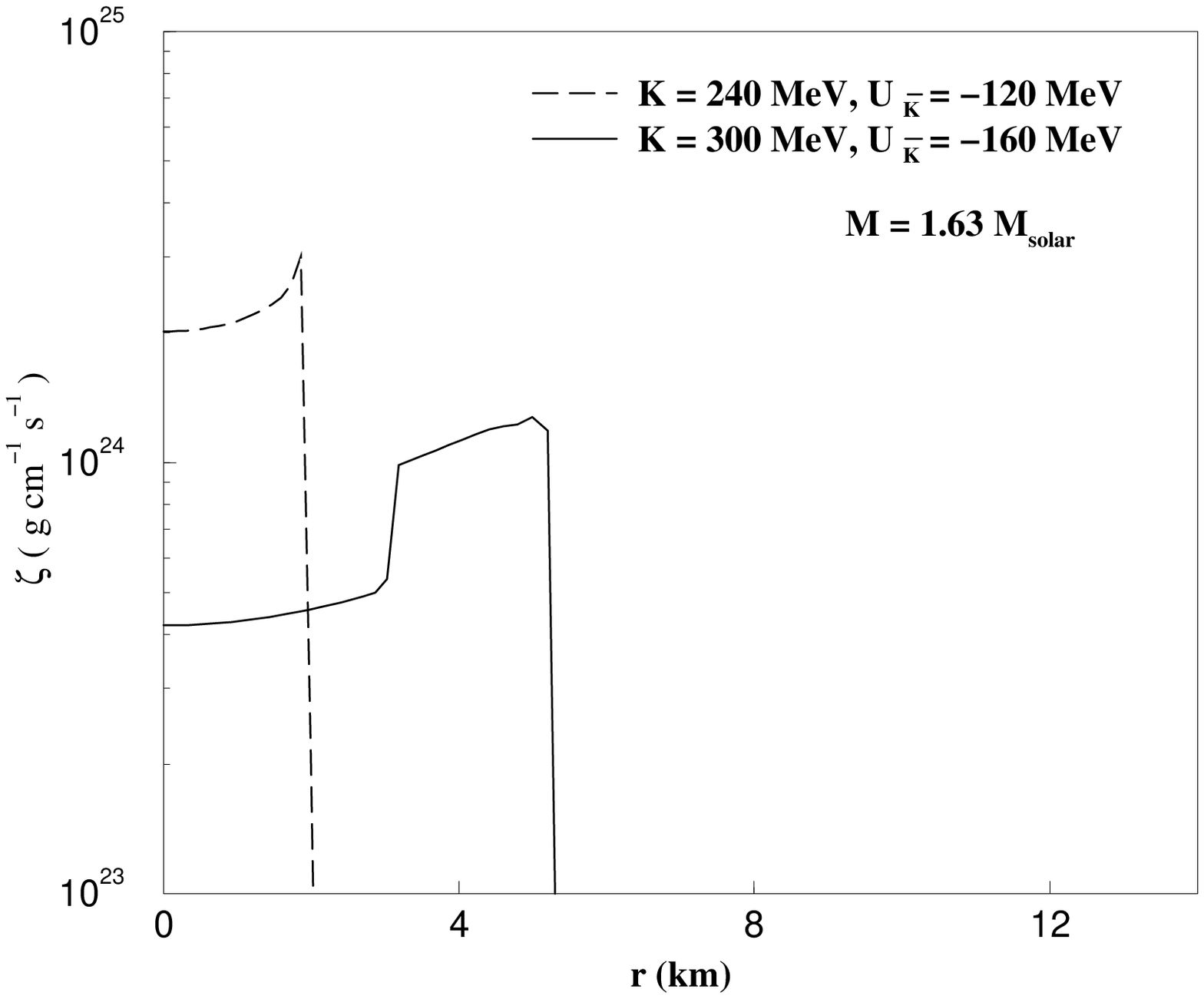}
}}

\vspace{4.0cm}

\noindent{\small{
Fig. 2. Bulk viscosity profile is plotted with equatorial distance for
a rotating neutron star of mass 1.63 M$_{\odot}$ for two parameter sets.}}

\newpage
\vspace{-2cm}

{\centerline{
\epsfxsize=14cm
\epsfysize=12cm
\epsffile{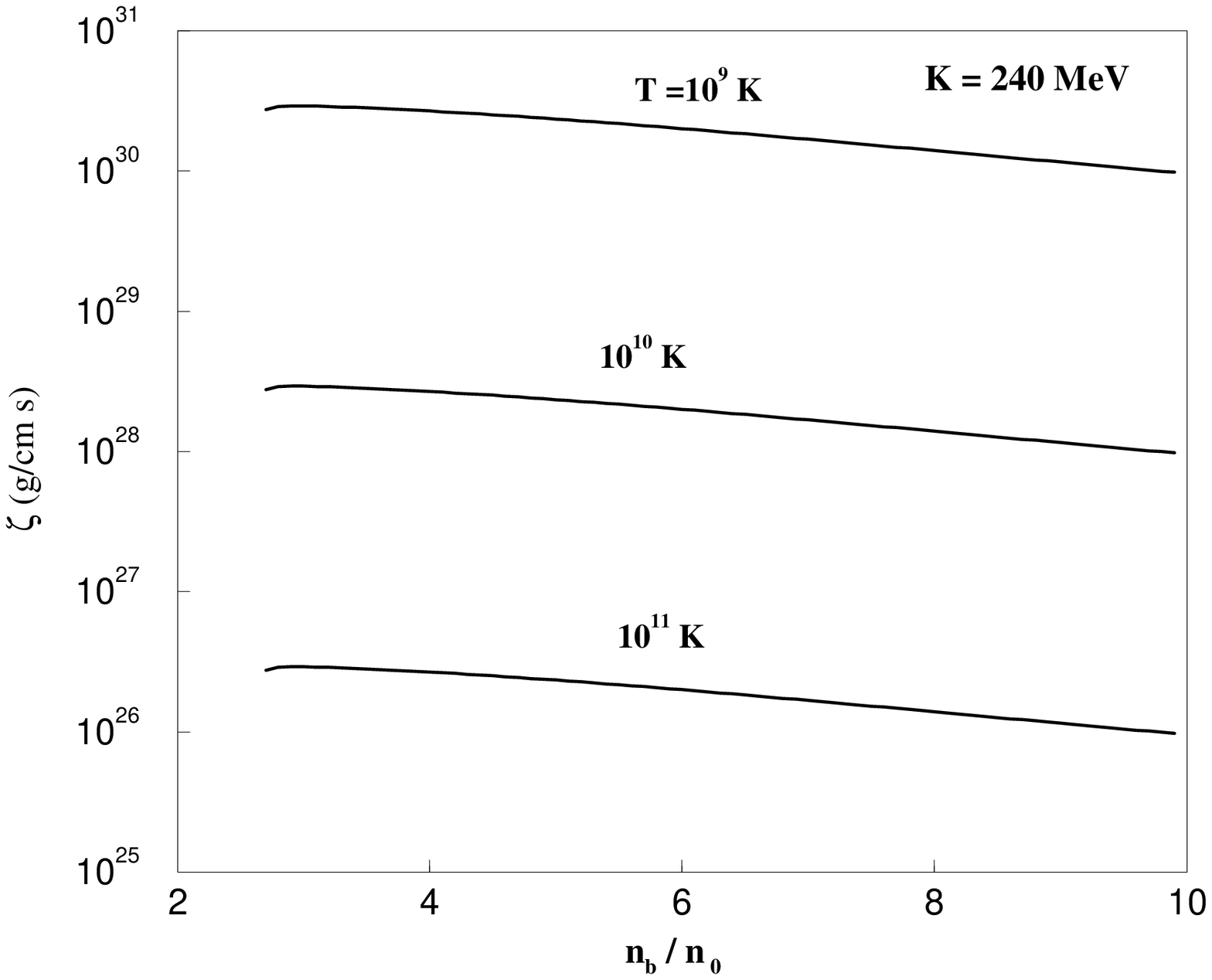}
}}

\vspace{4.0cm}

\noindent{\small{
Fig. 3. Hyperon bulk viscosity coefficient is exhibited as a function of normalised baryon density for different temperatures.}}

\newpage
\vspace{-2cm}

{\centerline{
\epsfxsize=14cm
\epsfysize=12cm
\epsffile{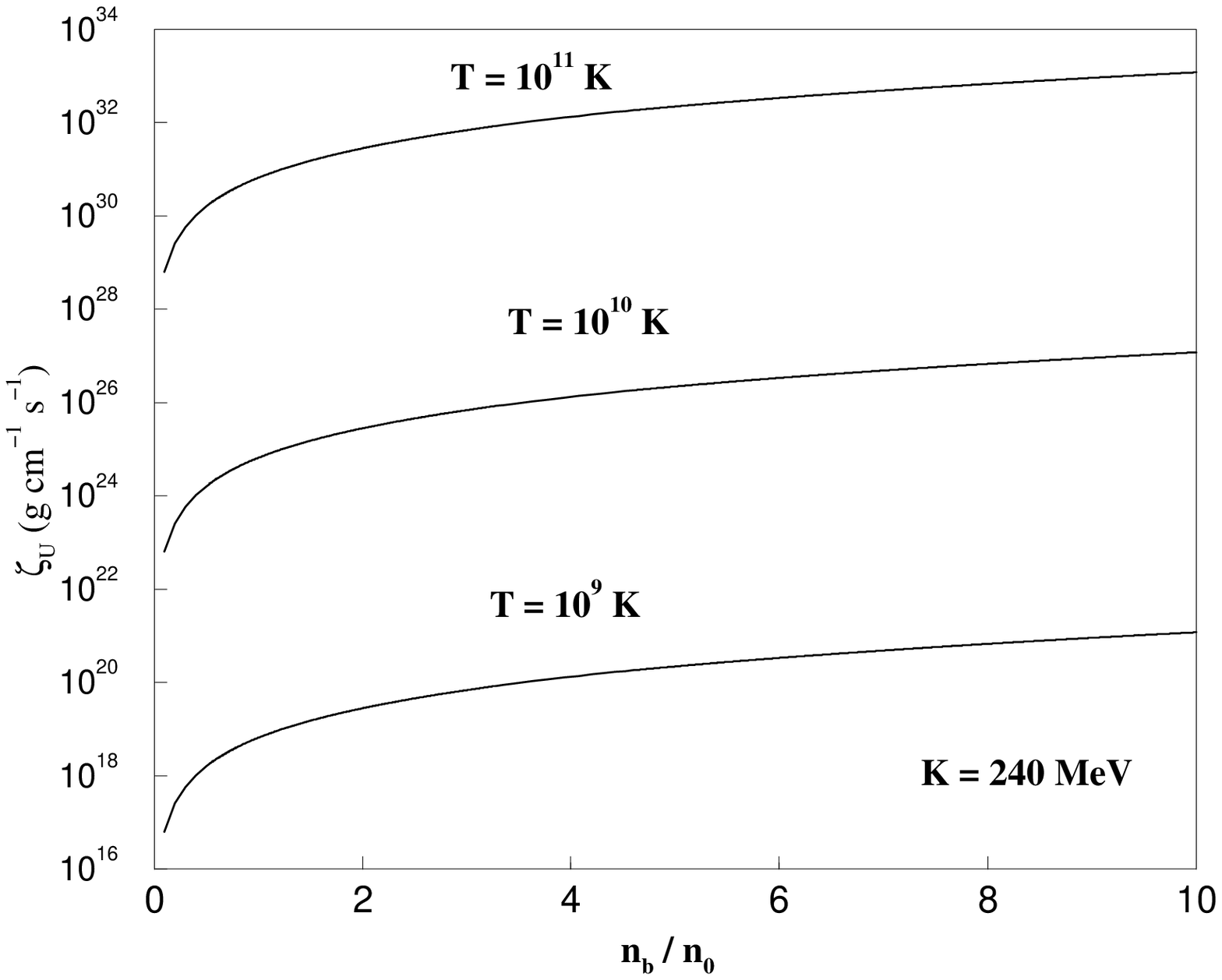}
}}

\vspace{4.0cm}

\noindent{\small{
Fig. 4. Density dependence of modified Urca bulk viscosity is shown for a range of temperatures.}}

\newpage
\vspace{-2cm}

{\centerline{
\epsfxsize=14cm
\epsfysize=12cm
\epsffile{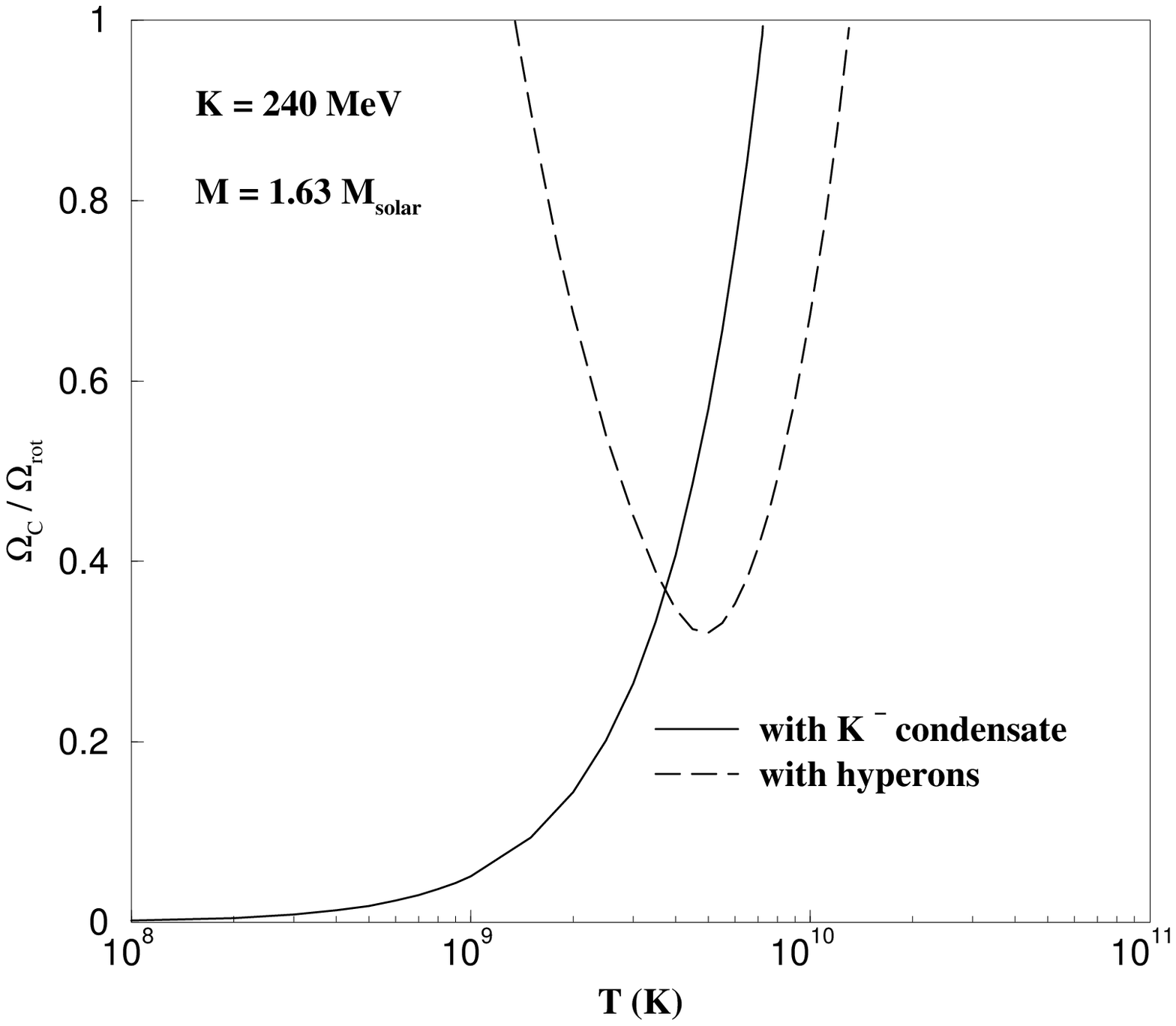}
}}

\vspace{4.0cm}

\noindent{\small{
Fig. 5. Critical angular velocities for 1.63 M$_{\odot}$ neutron star are
plotted as a function of temperature.}}


\begin{thebibliography}{40}
\bibitem {Jon1} P.B. Jones, Phys. Rev. Lett. {\bf 86}, 1384 (2001).
\bibitem {Jon2} P.B. Jones, Phys. Rev. D {\bf 64}, 084003 (2001).
\bibitem {Lin02} L. Lindblom and B.J. Owen, Phys. Rev. D {\bf 65}, 063006 
(2002).
\bibitem {Dal} E.N.E. van Dalen and A.E.L. Dieperink, Phys. Rev. C {\bf 69}, 
025802 (2004).
\bibitem {Dra} A. Drago, A. Lavagno and G. Pagliara, Phys. Rev. D {\bf 71}, 
103004 (2005).
\bibitem {Mad92} J. Madsen, Phys. Rev. D {\bf 46}, 3290 (1992).
\bibitem {Mad00} J. Madsen, Phys. Rev. Lett. {\bf 85}, 10 (2000).
\bibitem{Bog} J. Boguta and A.R. Bodmer, Nucl. Phys. {\bf A292}, 413 (1977).
\bibitem{Ser} B.D. Serot and J.D. Walecka, Adv. in Nucl. Phys. {\bf 16}, 1
(1986).
\bibitem{Gle99} N.K. Glendenning and J. Schaffner-Bielich, Phys. Rev.
C {\bf 60}, 025803 (1999).
\bibitem{Gle92} N.K. Glendenning, Phys. Rev. D {\bf 46}, 1274 (1992).
\bibitem {Lan} L.D. Landau and E.M. Lifshitz, Fluid Mechanics, 2nd ed.
(Butterworth-Heinemann, Oxford, 1999). 
\bibitem {And01} N. Andersson, Class. Quant. Grav. {\bf 20}, R105 (2003).
\bibitem {Mar} R.E. Marshak, Riazuddin and C.P. Ryan, Theory of weak 
interactions in particle physics (Wiley-Interscience, New York, 1969). 
\bibitem{Sch00} J. Schaffner-Bielich, R. Mattiello and H. Sorge, 
Phys. Rev. Lett. {\bf 84}, 4305 (2000).
\bibitem {Nar} M. Nayyar and B.J. Owen, Phys. Rev. D {\bf 73}, 084001 (2006).
\bibitem {Lin99} L. Lindblom, G. Mendell  and B.J. Owen, Phys. Rev. D {\bf 60},
064006 (1999).
\bibitem {Lin98} L. Lindblom, B.J. Owen and S. M. Morsink, Phys. Rev. Lett. 
{\bf 80}, 4843 (1998).
\bibitem {Saw} R.F. Sawyer, Phys. Rev. D {\bf 39}, 3804 (1989).
\bibitem{Gle91} N.K. Glendenning and S.A. Moszkowski, Phys. Rev. Lett. 
{\bf 67}, 2414 (1991).
\bibitem {Bani2} S. Banik and D. Bandyopadhyay, Phys. Rev. C {\bf 64}, 055805
(2001). 
\bibitem{Fri94} E. Friedman, A. Gal and C.J. Batty, Nucl. Phys. {\bf A579}, 
518 (1994);\\
C.J. Batty, E. Friedman and A. Gal, Phys. Rep. {\bf 287}, 385 (1997).  
\bibitem{Fri99} E. Friedman, A. Gal, J. Mare\v{s} and A. Ciepl\'y, 
Phys. Rev. C {\bf 60}, 024314 (1999).
\bibitem{Koc} V. Koch, Phys. Lett. B {\bf 337}, 7 (1994). 
\bibitem{Waa} T. Waas and W. Weise, Nucl. Phys.  {\bf A625}, 287 (1997).
\bibitem{Li} G.Q. Li, C.-H. Lee and G.E. Brown, Phys. Rev. Lett. 
{\bf 79}, 5214 (1997); Nucl. Phys. {\bf A625}, 372 (1997).
\bibitem{Pal2} S. Pal, C.M. Ko, Z. Lin and B. Zhang, Phy. Rev. C {\bf 62},
061903(R) (2000).
\bibitem{Sch} J. Schaffner and I.N. Mishustin, Phys. Rev. C {\bf 53}, 1416 
(1996).
\bibitem {Mis} J. Schaffner and I.N. Mishustin, Phys. Rev. C {\bf 53}, 1416 
(1996).
\bibitem {Dov} C. B. Dover and A. Gal, Prog. Part. Nucl. Phys. {\bf 12}, 171
(1984).
\bibitem {Sch94} J. Schaffner, C.B. Dover, A. Gal, D. J. Millener, C. Greiner 
and H. St\"ocker, Ann. Phys. (N.Y.) {\bf 235}, 35 (1994).
\bibitem {Chr} R. E. Chrien and C. B. Dover, Annu. Rev. Nucl. Part. Sci. 
{\bf 39}, 113 (1989).
\bibitem {Fuk} T. Fukuda et al., Phys. Rev. C {\bf 58}, 1306 (1998).
\bibitem {Kha} P. Khaustov et al., Phys. Rev. C {\bf 61}, 054603 (2000).
\bibitem {Fri} E. Friedman, A. Gal and C. J. Batty, Nucl. Phys. {\bf A579},
518 (1994); C. J. Batty, E. Friedman and A. Gal, Phys. Rep. {\bf 287}, 385
(1997).
\bibitem {Bart} S. Bart et al., Phys. Rev. Lett. {\bf 83}, 5238 (1999).
\bibitem {Sch93} J. Schaffner, C.B. Dover, A. Gal, C. Greiner and H. St\"ocker,
Phys. Rev. Lett. {\bf 71}, 1328 (1993).
\bibitem {Rati} D. Chatterjee and D. Bandyopadhyay, Phys. Rev. D {\bf 74}, 
023003 (2006); \\
D. Chatterjee and D. Bandyopadhyay, Astrophys. Space Sci. 308 (2007) 451;\\ 
\bibitem {Rati2} D. Chatterjee and D. Bandyopadhyay, Phys. Rev. D {\bf 75}, 
123006 (2007); \\
\end{thebibliography}
\end{document}